\documentclass[12pt]{article}
\usepackage[utf8]{inputenc}
\usepackage[english]{babel}
\usepackage{amssymb,graphicx}
\usepackage{amsmath,amsfonts}
\usepackage{fullpage}

\newcommand{\be}{\begin{equation}}
\newcommand{\ee}{\end{equation}}
\newcommand{\bea}{\begin{eqnarray}}
\newcommand{\eea}{\end{eqnarray}}

\title{\vspace{-2cm} 
\begin{flushright}
{\normalsize INR-TH-2019-003}
\end{flushright}
\vspace{0.5cm} 
Photon splitting constraint on Lorentz Invariance Violation from Crab Nebula spectrum}
\author{Konstantin Astapov, Dmitry Kirpichnikov,  Petr Satunin\thanks{{\bf e-mail}: satunin@ms2.inr.ac.ru}
\vspace{.2cm}\\
\normalsize\it  Institute for Nuclear Research of the Russian Academy
of Sciences, \\  
\normalsize \it  60th October Anniversary Prospect, 7a, 117312  Moscow, Russia} 
\date{}
\begin{document}
\maketitle

\begin{abstract}
We calculate the decay width of the photon splitting into three 
photons in a  model of quantum electrodynamics with broken Lorentz 
invariance. We show that this process can lead to a cut-off in 
the very-high-energy part of a photon spectra of astrophysical sources. We  
obtain the 95\% CL bound on the Lorentz violating mass scale for photons 
from the analysis of the very-high-energy part of the Crab Nebula spectrum, 
obtained by HEGRA. This bound improves previous constraints by more than an order of magnitude.
\end{abstract}

\section{Introduction}
\label{intro}

Lorentz invariance (LI) is one of the most fundamental principles of 
particle physics. However, in several theoretical models it can be broken. 
These models are mostly motivated by different approaches to construct the quantum theory of gravity (see review \cite{Liberati:2013xla} and references therein). According to some approaches \cite{Ellis:2003if, Horava:2009uw, Girelli:2012ju}, deviations from LI, tiny at laboratory energies, increase rapidly with energy and  
become large at a certain energy scale $M_{LV}$.

The most common approach is to consider Lorentz invariance violation (LV) in the matter sector  in the framework of effective field theory (EFT) \cite{Coleman:1997xq,Colladay:1998fq,Jacobson:2002hd,Myers:2003fd,
Kostelecky:2007fx}.  In this framework the existence of the preferred frame is assumed.
The generic effect of LV for particles is the modification of dispersion relations. At energies much smaller than the LV scale
$M_{LV}$ the dispersion relation for a certain particle can be expanded in powers of its momentum\footnote{The spacial isotropy in the preferred frame is assumed; otherwise the dispersion relation would depend both on the absolute value of the momenta and on its direction.}. Thus, for photons, we have: 
\begin{equation}
\label{IntroDisp}
E_\gamma^2=k_\gamma^2 + \sum_n \left( \pm \xi_n\right) \frac{k_\gamma^{n+2}}{M_{LV}^n}.
\end{equation}
where $\xi_n$ are parameters assumed to be of the order of unity, the sign of $\xi_n$ is shown explicitly\footnote{In some models the sign before a given LV term depend on photon polarization.}. The first non-zero  correction term 
in Eq.~(\ref{IntroDisp}) is the leading one. The commonly considered cases are $n=1$ (cubic term) and $n=2$ (quartic term). In the class of models, preserving CPT, the cubic term is absent, so the quartic term is the leading one. 

Possible violation of LI for photons may be tested in several ways. First, nontrivial photon dispersion may be tested in timing observations of variable distant sources \cite{Biller:1998hg}. Second, cross-sections of reactions involving particles with LV dispersion relation are modified \cite{Coleman:1997xq,Jacobson:2002hd,Colladay:2001wk,Rubtsov:2012kb}.   
Physical processes involving photons with modified dispersion relation~(\ref{IntroDisp}) can be classified by 
two kinematic types: the sign ``$+$'' is related to superluminal LV, sign ``$-$'' --- to subluminal case.
It is convenient to introduce effective photon mass, 
\begin{equation}
m^2_{\gamma,eff} \equiv E_\gamma^2-k_\gamma^2,
\label{EffectiveMass}
\end{equation}
 that is associated with 
superluminal and subluminal photons for $m^2_{\gamma,eff}>0$ and 
$m^2_{\gamma,eff}<0$ respectively.
The processes, constraining the subluminal LV such as pair production on soft photon background\cite{Kifune:1999ex,Stecker:2001vb,Stecker:2003pw,Jacob:2008gj,Tavecchio:2015rfa,Abdalla:2019krx} and in Coulomb field\cite{Vankov:2002gt,Rubtsov:2012kb,Rubtsov:2016bea} have been discussed elsewhere. In the present paper we focus on the superluminal case. 

Important constraints in the superluminal case come from the photon decay to electron-positron pair $\gamma \to e^+ e^-$. 
 Indeed, energy-momentum conservation  implies that this process becomes allowed if the photon effective mass  exceeds the double electron 
mass \cite{Coleman:1997xq}, $m_{\gamma,eff} \geq 2m_e$.  The equality determines an 
 energy threshold for this reaction, which means that the 
photons with energy lower than the 
threshold do not decay.  For energies larger than the threshold, 
the photon decay occurs very rapidly 
\cite{Rubtsov:2012kb}.

However, even if the decay $\gamma \to e^+e^-$ is forbidden, a photon with superluminal dispersion relation is not a stable particle. The corresponding decay channel is photon splitting to several photons, $\gamma \to N\gamma$.
In the context of LV, photon splitting was considered  in quantum electrodynamics (QED) with additional Chern-Simons term in \cite{Adam:2002rg}, and for the case of cubic photon dispersion relation in \cite{Gelmini:2005gy} ($n=1$ in (\ref{IntroDisp})).

In the standard Lorentz Invariant QED the process of photon 
splitting does not occur\footnote{Photon splitting process 
$\gamma  \to \gamma\gamma$ occurs in LI QED in an external 
magnetic field \cite{Adler:1971wn,BialynickaBirula:1970vy}.}  due to the energy-momentum 
conservation as all outgoing photon momenta must be parallel. In this 
configuration the phase volume, as well as the amplitude, is equal 
to zero. 

In the presence of LV the photon dispersion relation is modified 
(\ref{IntroDisp}), and the kinematical configuration for splitting, 
determined by the energy-momentum conservation, changes. The key 
feature of dispersion relation (\ref{IntroDisp}) for splitting is 
that the photon effective mass  (\ref{EffectiveMass}) depends on its 
energy faster than linearly. In this case the energy-momentum 
conservation  implies that the sum of effective masses for outgoing 
photons is less than the effective mass for the initial photon 
$\sum_i m^{out}_{eff, \, i} < m_{eff}^{in}$.  The photon splitting 
has no threshold: it is kinematically allowed whenever the 
dispersion relation (\ref{IntroDisp}) is superluminal. However, the 
rate of photon splitting is significantly suppressed compared to the 
rate of photon decay. The main reason is  that the corresponding 
process has a small phase volume  of the outgoing particles.

At first glance it seems that the photon splitting into two 
photons, $\gamma \to 2 \gamma$, is the leading order splitting process. In LI QED the corresponding matrix element is exactly zero due to the cancellation of fermionic and anti-fermionic loops, which is the statement of the Furry theorem. 
If CPT-parity in the fermion sector of the model is broken, the Furry theorem does not hold anymore.  However, even in this case the splitting width acquires additional suppression factor, see a discussion in~\cite{Liberati:2013xla}. Moreover, if the model contains photons with only two transverse polarizations as in the standard case, the splitting channel $\gamma \to 2\gamma$ is forbidden due to helicity conservation.

Thus, the main  splitting process is the photon decay 
 into 3 photons,  $\gamma \to 3\gamma$. 
 The estimation for the decay width of this process in terms of the 
photon effective mass (\ref{EffectiveMass}) was done in Ref. 
\cite{Gelmini:2005gy}. The authors  treated the initial photon as a massive 
particle with the effective mass defined by (\ref{EffectiveMass}), while 
three outgoing photons were considered to be massless. First, the width of the splitting process was estimated in an artificial ``rest frame'' of the initial photon. Dimensionless phase space integral of the matrix element has not been computed explicitly but considered as a parameter of the order of unity. Second, the authors of Ref.~\cite{Gelmini:2005gy}  made a boost into the laboratory frame, multiplying the splitting width to the inverse gamma-factor, $m_{\gamma,eff}/E_\gamma$. Thus, the estimation for the splitting rate~\cite{Gelmini:2005gy} reads,
\begin{equation}
\label{Gelmini}
\Gamma_{\gamma\to 3\gamma} \simeq \left(\frac{2\alpha^2}{45}\right)^2 \frac{1}{3!2^{11}\pi^9}\frac{m_{\gamma, eff}^{10}}{m_e^8 E_\gamma}\,\times\, f,
\end{equation}
where $m^2_\gamma$ is determined by (\ref{EffectiveMass}) and $f$ is a dimensionless phase volume which has not been computed in \cite{Gelmini:2005gy}. 
Thus, the estimation of \cite{Gelmini:2005gy} is order of magnitude,  so that more 
accurate calculation of the splitting rate seems to be necessary. 
   In our paper we perform direct calculation of the photon 
  splitting width in the laboratory frame, integrating properly the
   phase space volume of the outgoing photons. In addition, we obtain momentum distribution for the splitting products which cannot be properly obtained by the estimation of \cite{Gelmini:2005gy}.

The article is organized as follows. In Sec.2 we describe the model and Feynman rules. In Sec.3 we calculate the matrix element of the splitting process and integrate it over the phase volume. Sec.4 is devoted to establishing constraints on LV mass scale from Crab nebula spectrum, in Sec. 5 we present discussion.

\section{The model.}

In order to perform direct calculation one must start from the Lagrangian formalism. The Lagrangian, appropriate to the dispersion relation (\ref{intro}) with $n=2$ and sign $+$ before LV term, is
\begin{equation}
\label{L1}
\mathcal{L}=-\frac{1}{4}F_{\mu\nu}F^{\mu\nu}  - \frac{1}{2 M_{LV}^2}F_{ij}\Delta^2 F^{ij}  +i\bar{\psi}\gamma^\mu D_\mu\psi - m\bar{\psi}\psi, \quad\qquad E_\gamma^2 = k_\gamma^2 + \frac{k_\gamma^4}{M_{LV}^2}.
\end{equation}
Here we rewrote the dispersion relation for a photon as well. One may also consider in Lagrangian (\ref{L1}) terms, violating LI for electrons. However, we will  omit them for the following reason.   Indeed, the bound on LV mass scale for electron $M_{LV,\,e}$  is set at the level of $10^{16}\,\mbox{GeV}$  in~\cite{Liberati:2012jf}, while the current limits \cite{Vasileiou:2013vra, Martinez-Huerta:2016azo}  on $M_{LV}$ for photons are of the order of $10^{11}-10^{12}\,\mbox{GeV}$. 

The model (\ref{L1}) corresponds to the most general model called non-minimal SME 
\cite{Kostelecky:2009zp} with a single nonzero coefficient $c^{(6)}_{(I)00} = - \sqrt{\pi}/M_{LV}^{2}$.
On the other hand, the model (\ref{L1}) coincides with  that of \cite{Rubtsov:2012kb}, in which fermions are set to be LI. All Feynman rules necessary for perturbative calculation in this model were obtained in \cite{Rubtsov:2012kb}. 
In particular, polarization sums for photons\footnote{Gauge invariance  still holds so photon still has only two physical polarizations.} can be written as
\begin{equation}
\label{polarizationsums}
\sum_{s=1,2}\varepsilon_\mu^{*(s)}(k_\gamma)\varepsilon^{(s)}_\nu(k_\gamma) = -g_{\mu\nu} - \frac{E_\gamma^2}{M_{LV}^2}u_\mu u_\nu,
\end{equation}
where $u_\mu=(1,0,0,0)$ is unit timelike 4-vector. In the LI limit $M_{LV} \to \infty$ the standard result is restored.

Integrating out electrons produces an effective four-photon vertex that can be read off the Euler-Heisenberg Lagrangian, 
\begin{equation}
\mathcal{L}_{E\mbox{-}H}=\frac{2 \alpha^2}{45 m_e^4}
\left[\left(\frac 12 F_{\mu\nu}F^{\mu\nu}\right)^2 
+7 \left(\frac 18 \epsilon^{\mu\nu\rho\sigma} 
F_{\mu\nu}F_{\rho\sigma}\right)^2\right].
\label{euler}
\end{equation}
 This form of Euler-Heisenberg Lagrangian is applicable in the limit $m_{\gamma, eff} \ll m_e$ -- far from the threshold of photon decay. 

Let us provide the current photon decay limit for model (\ref{intro}). Photon effective mass is expressed as 
$m^2_{\gamma,eff} =\frac{E_\gamma^4}{M_{LV}^2}$.
Hence, detection of a photon with energy $E_\gamma$ establishes the constraint 
$M_{LV} > \frac{E_\gamma^2}{2m_e}\;$.
The recent analysis~\cite{Martinez-Huerta:2016azo} 
using the highest-energy photons observed from
the Crab nebula sets the constraint,
\begin{equation}
\label{decayrecent}
M_{LV} > 2.8\times 10^{12}\,\mbox{GeV}.
\end{equation}

Let us compare the constraint (\ref{decayrecent}) with the estimation for 
photon splitting constraint from Ref.~\cite{Gelmini:2005gy}. Although in Ref.~\cite{Gelmini:2005gy} the numerical estimations for the constraint were performed only for the cubic dispersion relation, $n=1$, in the formalism of effective masses (see  (\ref{Gelmini})) they can be easily transferred to the quartic, $n=2$.  Thus, using the fact that the photon with energy 
of $50$ TeV have not been splitted during their propagation from Crab nebula, and the estimation (\ref{Gelmini}) with quartic photon dispersion relation (\ref{L1}), we can estimate  the bound on $M_{LV}$ of the order of $10^{13}$ GeV. 
This  expected limit is order of magnitude higher
 than the bound from the absence 
of the photon decay (\ref{decayrecent}). Therefore, the 
splitting process 
is relevant for setting a more stringent
 constraint on $M_{LV}$ than the process of decay to $e^+e^-$ pair. 
  So that, it is instructive to perform accurate direct calculation of the splitting 
 width in the laboratory frame, and set more precise bound on $M_{LV}$.  The latter is the focus of the  present paper.

\section{Decay width}
 In this section we calculate the  width
 of the photon splitting  in the model (\ref{intro}), using 
  Euler-Heisenberg Lagrangian (\ref{euler}) for four-photon
  interaction. 
Let us first consider the matrix element  of the process 
$\gamma \rightarrow 3 \gamma$.

\subsection{Matrix element}

 Given the Euler-Heisenberg Lagrangian (\ref{euler}),  one  can extract the  matrix element of photon splitting process in a straightforward way. The lowest order amplitude for four-photon process can be written as
\begin{align}
\label{MatrElement1}
\mathcal{M} = \frac{\alpha^2}{90 m_e^4}\left[(k^1_\mu \varepsilon_\nu^1 - k^1_\nu \varepsilon_\mu^1)(k^2_\mu \varepsilon_\nu^2 - k^2_\nu \varepsilon_\mu^2)(k^3_\rho \varepsilon_\lambda^3 - k^3_\lambda \varepsilon_\rho^3)(k_\rho \varepsilon_\lambda^{*} - k_\lambda \varepsilon_\rho^{*})\right. + \notag \\
+ \frac{7}{16}\left[\epsilon^{\mu\nu\rho\lambda}(k^1_\mu \varepsilon_\nu^1 - k^1_\nu \varepsilon_\mu^1)(k^2_\rho \varepsilon_\lambda^2 - k^2_\lambda \varepsilon_\rho^2) \right]\left[\epsilon^{\alpha\beta\gamma\delta}(k^3_\alpha \varepsilon_\beta^3 - k^3_\beta \varepsilon_\alpha^3)(k_\gamma \varepsilon_\delta^{*} - k_\delta \varepsilon_\gamma^{*}) \right] + \\
+ \left.\mbox{permutations}\;\right], \ \ \ \notag  
\end{align}
here $k^i_\mu, \ \ i=1,2,3$ are outgoing photon four-momenta, $k_\mu$ is four-momentum for initial photon,  $\varepsilon^i_\mu$ and $ \varepsilon_\mu$ 
 are the  relevant polarization vectors.  Note that photon splitting, $\gamma \rightarrow 3 \gamma$, is a cross-channel of a well-known process of  two photon scattering, $ \gamma \gamma \rightarrow \gamma \gamma$.  Therefore, the matrix element  of interest has the same structure  up to the initial momentum redefinition $k \to -k$ (compare  with the matrix element for the photon scattering in \cite{Schwartz}). 
However there are two typical differences between the LI and LV cases that must be taken into account.
First, the squared  four-momenta of external on-shell photons are not
equal zero in LV case. 
 Second, the polarization sums of photon are modified by (\ref{polarizationsums}).

  The evaluation of matrix element (\ref{MatrElement1}) by taking 
 into account polarization 
 sums (\ref{polarizationsums}) and all permutations  is somewhat 
 lengthy,   so that  we apply the following technique for 
 simplifications.
 Namely, we factorize polarization vectors in the matrix element 
 to apply  the modified polarization sums  (\ref{polarizationsums})
  in a straightforward way. Thus, one has 
\begin{equation}
\label{MatrixElementTensor}
\mathcal{M} = \mathcal{M}_{\mu\nu\rho\lambda}(k_1,k_2,k_3,k) \varepsilon_\mu(k_1) \varepsilon_\nu (k_2)\varepsilon_\rho(k_3) \varepsilon_\lambda^*(k).
    \end{equation}
The matrix element tensor $\mathcal{M}_{\mu\nu\rho\lambda}$, which  is
 independent on polarization vectors,  can be splitted to a normal and dual parts.  Such decomposition can be written as
\begin{equation}
\label{MatrElementTwoParts}
\mathcal{M}_{\lambda_1\lambda_2\lambda_3\lambda}= A_{\lambda_1\lambda_2\lambda_3\lambda} + \frac{7}{16} \tilde{A}_{\lambda_1\lambda_2\lambda_3\lambda}.
\end{equation}
Each term in (\ref{MatrElementTwoParts})
 is expressed in the following way:
\begin{align}
\label{MatrElementNormalDual}
  A_{\lambda_1\lambda_2\lambda_3\lambda}=8(T_{\lambda_1\lambda_2}(k_1,k_2)T_{\lambda_3\lambda}(k_3,k) + T_{\lambda_1\lambda_3}(k_1,k_3)T_{\lambda_2\lambda}(k_2,k) + T_{\lambda_1\lambda}(k_1,k)T_{\lambda_3\lambda_2}(k_3,k_2) )
, \notag \\
  \tilde{A}_{\lambda_1\lambda_2\lambda_3\lambda}=8(\tilde{T}_{\lambda_1\lambda_2}(k_1,k_2)\tilde{T}_{\lambda_3\lambda}(k_3,k) + \tilde{T}_{\lambda_1\lambda_3}(k_1,k_3)\tilde{T}_{\lambda_2\lambda}(k_2,k) + \tilde{T}_{\lambda_1\lambda}(k_1,k)\tilde{T}_{\lambda_3\lambda_2}(k_3,k_2) ).
\end{align}  
Here we introduced tensors
\begin{equation}
\label{TMuNu}
    T_{\mu\nu}(k,p) = 2 (pk)g_{\mu\nu} - 2 p_\mu k_\nu, \qquad \ 
    \tilde{T}_{\mu\nu}(k,p) = - 4 k^{\rho}p^{\lambda} \varepsilon^{\mu\nu\rho\lambda}.
\end{equation}
Calculating the matrix element (\ref{MatrixElementTensor}) squared, we average over initial photon polarization, and sum over final photon polarizations, 
\begin{align}
\label{SqMatrixElement} 
\overline{\left| \mathcal{M}\right|^2}=\frac{1}{2} \sum_{pols} \mathcal{M}^*\mathcal{M} =  \mathcal{M}^*_{\alpha\beta\gamma\delta} \mathcal{M}_{\mu\nu\rho\lambda}\;& \sum_{s_1} \varepsilon^{*(s_1)}_\alpha(k_1)\varepsilon^{(s_1)}_\mu(k_1) \;\sum_{s_2} \varepsilon^{*(s_2)}_\beta (k_2) \varepsilon^{(s_2)}_\nu (k_2)\; \cdot \\ \cdot &\sum_{s_3} \varepsilon^{*(s_3)}_\gamma(k_3)\varepsilon^{(s_3)}_\rho(k_3) \;\frac{1}{2} \sum_{s} \varepsilon^{(s)}_\delta(k)\varepsilon^{*(s)}_\lambda(k).  \notag
\end{align}
 Note that the  polarization sums 
$\sum_{s} \varepsilon^{*(s)}_\mu(k)\varepsilon^{(s)}_\nu(k)$ are given by 
Eq.~(\ref{polarizationsums}), the matrix element tensor 
$\mathcal{M}_{\mu\nu\rho\lambda}$ is determined by eqs. 
(\ref{MatrElementTwoParts})-(\ref{TMuNu}). The calculation 
(\ref{SqMatrixElement}) was done via FeynCalc package 
\cite{Mertig:1990an,Shtabovenko:2016sxi} for Wolfram Mathematica. The result 
for the squared matrix element cannot be given in a short formula and is 
shown elsewhere \cite{GITHUB}. The squared matrix element 
(\ref{SqMatrixElement}), constrained to LI theory and applied to the cross-channel of splitting, 
photon scattering $\gamma\gamma \to \gamma\gamma$, agrees with the standard result 
\cite{Schwartz}.

\subsection{Phase volume integration}
In this subsection we discuss the integration of the squared matrix 
element (\ref{SqMatrixElement}) over 3-particle phase volume in order to 
calculate  the photon splitting decay width. The general textbook formula for the decay width reads
\begin{equation}
\label{PhaseVolume_1}
\Gamma_{\gamma\to 3\gamma} =  \frac{1}{2E}\int \frac{1}{3!} \prod_{i=1,2,3}\frac{d\vec{k}_i}{(2\pi)^3(2E_i)}\overline{|\mathcal{M}|^2} (2\pi)^4\,\delta^{(4)}\left(k -k_1-k_2 - k_3\right),
\end{equation}
where  $E_i$ are the energies  of  outgoing photons $ (i=1,2,3)$, and 
$E$ is the energy of initial photon, which is considered to be small compared to $M_{LV}$, $E \ll M_{LV}$. The factor $1/3!$ 
in (\ref{PhaseVolume_1}) accounts the  permutation   
number of the final photons.

Since our theory does not possess LI, we are not allowed to make a boost to a center-of-mass frame for simplicity. Hence, we have to carry out the calculations in the laboratory frame.
We perform integration over final momenta in (\ref{PhaseVolume_1}) in cylindric coordinates, dividing the spacial momenta of final 
photons to longitudinal $k_i^\parallel$ and transverse 
$\vec{k}_i^\perp$ components with respect to the initial photon momentum $k$.
Transverse components $\vec{k}_i^\perp$, in turn, are 2-vectors in a plane, perpendicular to $\vec{k}$. Each of them can be parametrized by the absolute value $|\vec{k}_i^\perp|$ and polar angle $\varphi_i$. One of the polar angles, say $\varphi_1$, is a free parameter corresponding to the rotational symmetry around initial photon momentum axis. The momentum conservation law implies that 
$\vec{k}_1^\perp + \vec{k}_2^\perp +\vec{k}_3^\perp = 0$. In terms of absolute values and polar angles, we have:
\begin{equation}
\label{Perp}
|\vec{k}_3^\perp|^2 = |\vec{k}_1^\perp|^2 + |\vec{k}_2^\perp|^2 + 2
|\vec{k}_1^\perp| |\vec{k}_2^\perp| \cos \varphi_{12},
\end{equation}  
$\varphi_{12}$ denotes the angle between $\vec{k}_1^\perp$ and $\vec{k}_2^\perp$. The conservation law for longitudinal momenta implies that  $k_1^\parallel +k_2^\parallel + k_3^\parallel = k$.


Note that  the phase volume integral (\ref{PhaseVolume_1})  vanishes  in LI theory since for LI photon dispersion relation $E_i^2=|\vec{k}^2_i|$  the delta-functions for energies and longitudinal momenta in (\ref{PhaseVolume_1}) coincide, fixing 
all transverse components $\vec{k}_i^\perp$ to be exactly zero.
In the presence of LV for photons nonzero transverse momenta are allowed. 
We integrate out 3-momentum $\vec{k}_3$ in (\ref{PhaseVolume_1}) eliminating spacial part of the delta-function, so the phase volume (\ref{PhaseVolume_1})  reads as
\begin{equation}
\label{PhaseVolume_2}
\Gamma_{\gamma\to 3\gamma} = \frac{1}{2^9\,3!\,\pi^5}\frac{1}{E}\int \frac{dk^\parallel_1\, dk^\parallel_2\, d^2k_1^\perp d^2k_2^\perp}{E_1 E_2 E_3} \delta(E-E_1-E_2-E_3)\overline{|\mathcal{M}|^2}.
\end{equation}
In order to perform integration in (\ref{PhaseVolume_2}) we should first resolve final energies $E_i$ through longitudinal and transverse momenta. Thus,
\begin{equation}
E_i^2= (k^\parallel_i)^2 + (k_i^\perp)^2 + \frac{\left((k^\parallel_i)^2 + (k_i^\perp)^2\right)^2}{M_{LV}^2}, \qquad i=1,2,3,
\end{equation}
while $E^2=k^2+\frac{k^4}{M_{LV}^2}$, $k=\sum_i k_i^\parallel$. Note that generally the squared transverse momenta are of the order of squared effective mass for initial photon but should not exceed it, $(k_i^\perp)^2 \lesssim \frac{k^4}{M_{LV}^2}$. Thus, for further calculations we apply the following approximation:
\begin{equation}
k_i^\perp / k \lesssim k/M_{LV} \ll 1.
\end{equation}
We perform our calculations in the leading order in this small parameter. 
Note that this parameter is associated with typical angle between 
the momenta of outgoing photons and the initial photon momentum.
It is convenient to express the
 longitudinal $k_i^\parallel$ 
  and transverse $|\vec{k}_i^\perp|$ momenta via dimensionless 
  variables  
   $\alpha_i$ and $\beta_i$ 
\begin{equation}
\label{Beta}
k_i^\parallel  = \alpha_i k, \qquad |\vec{k}_i^\perp| = \frac{k^2}{M_{LV}}\cdot \beta_i, \qquad i=1,2,3. 
\end{equation} 
In this notations, energies for outgoing photons, including the first order correction, take the form
\begin{equation}
\label{E12}
E_i=k\alpha_i + \frac{k^3}{2M_{LV}^2} \left(\alpha^3_i + \frac{\beta^2_i}{\alpha_i}\right), \ \ i=1,2,
\end{equation}
\begin{equation}
\label{E3}
E_{3} = k\,(1-\alpha_1 - \alpha_2) +\frac{k^3}{2 M_{LV}^2}\left((1-\alpha_1 - \alpha_2)^3 + \frac{\beta_1^2 +\beta_2^2 +2\beta_1\beta_2\cos\varphi_{12}}{1-\alpha_1 - \alpha_2}\right).
\end{equation}
The expressions (\ref{Beta})-(\ref{E3}) should be substituted to the delta-function in (\ref{PhaseVolume_2}).  We are interested in the leading order term, so we neglect the LV corrections to energies in the denominator of (\ref{PhaseVolume_2}).
Then the phase volume (\ref{PhaseVolume_2}) reads
\begin{align}
\label{PhaseVolume_3}
&\Gamma_{\gamma\to 3\gamma} \; =\; \frac{1}{2^9\,3!\,\pi^5}\frac{2\pi}{k} \left( \frac{k^2}{M_{LV}}\right)^4 \int \frac{d\alpha_1\, d\alpha_2\, d\varphi_{12} d(\beta_1^2) d(\beta_2^2)}{\alpha_1\alpha_2(1-\alpha_1-\alpha_2)}\ \overline{|\mathcal{M}|^2}\ \times \\
\times\;  \delta&\left[\frac{k^3}{2M_{LV}^2} \left( 3(\alpha_1+\alpha_2)(1-\alpha_1-\alpha_2 + \alpha_1\alpha_2) + \frac{\left( \frac{1-\alpha_2}{\alpha_1}\beta_1^2 + \frac{1-\alpha_1}{\alpha_2}\beta_2^2 + 2\beta_1 \beta_2 \cos \varphi_{12} \right)}{1-\alpha_1-\alpha_2} \right)   \right]. \notag
\end{align} 
Here additional $2\pi$ factor is the result of integration over 
$\varphi_1$. The delta-function in  (\ref{PhaseVolume_3}) can be 
removed by taking the integral over $\varphi_{12}$.
 In what follows we express $\cos \varphi_{12}$ as the function of 
 dimensionless variables
\begin{equation}
\label{cos_phi_2}
\cos \varphi_{12}=\frac{3(\alpha_1+\alpha_2)(1-\alpha_1-\alpha_2)(1-\alpha_1-\alpha_2 + \alpha_1\alpha_2) -\frac{1-\alpha_2}{\alpha_1}\beta_1^2 - \frac{1-\alpha_1}{\alpha_2}\beta_2^2  }{2\beta_1\beta_2}.
\end{equation}
Note that the decay width (\ref{PhaseVolume_3}) acquires 
 additional factor $\frac{k^3}{2M_{LV}^2}\frac{2\beta_1\beta_2}{1-\alpha_1\alpha_2}\sin\varphi_{12}$ in the denominator.  Finally, the  decay width  reads as
\begin{equation}
\label{PhaseVolume_4}
\Gamma_{\gamma\to 3\gamma} = \frac{1}{2^8\,3!\,\pi^4}\frac{k^4}{M_{LV}^2}\int \frac{d\alpha_1\, d\alpha_2\, d\beta_1 d\beta_2}{\alpha_1\alpha_2} \frac{\overline{|\mathcal{M}|^2}}{\left. \sin\varphi_{12} \right|_{\varphi_{12}=\varphi_{12}(\alpha_1,\alpha_2,\beta_1,\beta_2)}}.
\end{equation}
Let us specify the  integration area for the variables in the phase volume. First, in our approach
all longitudinal parts of outgoing photon momenta should be positive, 
hence, 
the area of integration over $\alpha_1,\alpha_2$ is a triangle: 
$\alpha_1>0,\,\alpha_2>0,\, \alpha_1+\alpha_2<1$. Second, the integration area 
 spanned by
$(\beta_1,\; \beta_2)$ is determined by the condition (see (\ref{cos_phi_2})): 
\begin{equation}
-1<\left.\cos \varphi_2\right|_{\varphi_2=\varphi_2(\alpha_1,\alpha_2,\beta_1,\beta_2)} <1.
\label{cosreal}
\end{equation}

\begin{figure}[t!]
\centering
\hspace{-4mm}
\includegraphics[width=0.5\linewidth]{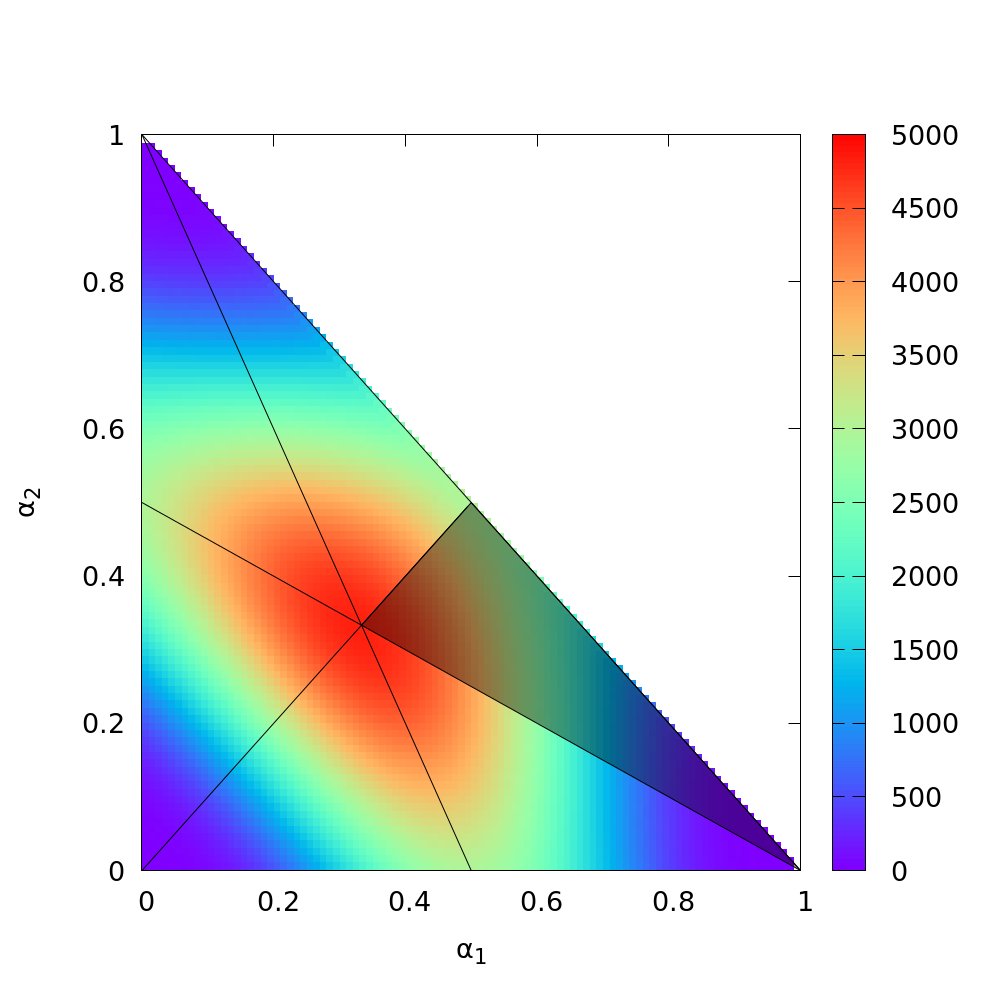}\hspace{-4mm}
\includegraphics[width=0.53\linewidth]{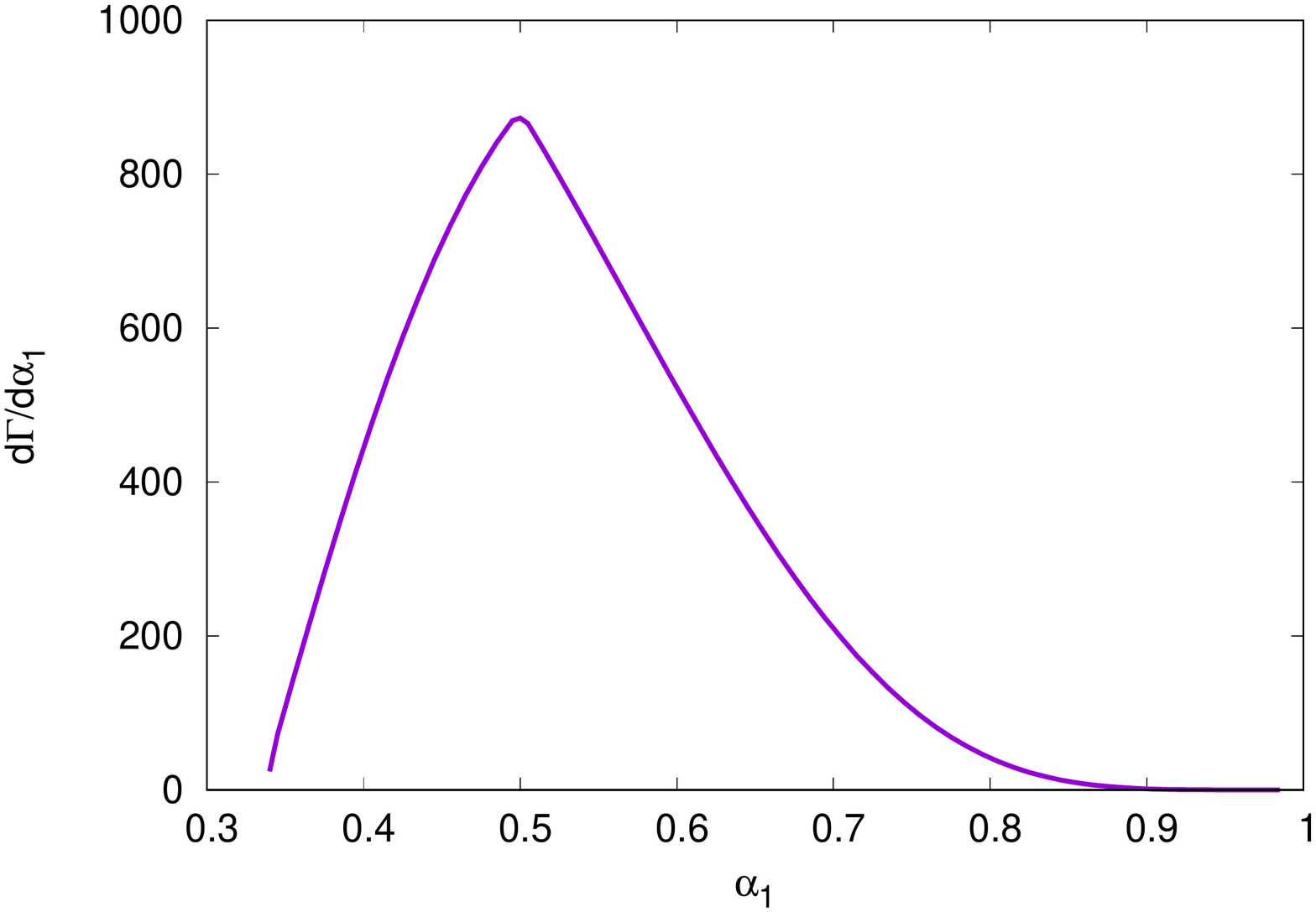}\hspace{-6mm}
\caption{Left panel: The decay width density $d^2\Gamma/d\alpha_1 
d\alpha_2$ as the function of longitudinal momenta of two outgoing photons 
$(\alpha_1,\alpha_2)$. The maximum is achieved  at $\alpha_1=\alpha_2=1/3$.
Shaded region corresponds to the hierarchy $k_1>k_2>k_3$, the width density 
in other regions is determined through the shaded region via the 
permutation symmetry. Right panel: The decay width density 
$d\Gamma/d\alpha_1$ as the function of the maximal longitudinal 
momentum $  \alpha_1 \equiv \alpha_{max}$. The peak is at $\alpha_1=0.5$. }
\label{Density}
\end{figure}

 We perform the integration in (\ref{PhaseVolume_4}) numerically. The 
matrix element (\ref{SqMatrixElement}) is evaluated at our 
kinematical configuration (the explicit result is lengthy and presented in \cite{GITHUB}). 
Fixing longitudinal momenta $(\alpha_1,\alpha_2)$, we first perform the integration
over transverse parts $(\beta_1,\; \beta_2)$.
The splitting width density distribution in terms of longitudinal momenta 
$\alpha_1,\,\alpha_2$ is shown in left panel of Fig.~\ref{Density}.
The relevant density distribution has a peak  at  
$\alpha_1 = \alpha_2 = 1/3$, that is the most probable collinear momenta
orientation of the outgoing photons, such that $\alpha_3=1/3$. On the other hand, for the fixed 
$\alpha_1 \sim 0$ it has local peak at $\alpha_2 =1/2$. This typical momentum 
orientation describes the collinearity of two out of three outgoing photons
$\alpha_2=\alpha_3=1/2$ with negligibly small momentum of a third photon
$\alpha_1=0$.
Remarkably, that the latter momentum configuration  is not 
in fact suppressed in comparison with the symmetric configuration
$\alpha_1=\alpha_2=\alpha_3=1/3$.
 Indeed, it corresponds to the phase volume of $\gamma \to \gamma\gamma$ which is 
not zero. In addition, we note that the probability of processes with one of final photons carrying almost all 
of the initial energy, say $\alpha_1\sim \alpha_2\sim 0$, $\alpha_3\sim 1$, 
appears to be strongly suppressed.

One should also note that the symmetry of permutations for final photons 
$\alpha_i \leftrightarrow \alpha_j, \ i,j=1,2,3,\ i \neq j, \ \alpha_3=1-\alpha_1-\alpha_2$, 
technically broken in our calculations, 
has been restored now. In the left panel of Fig.~\ref{Density} we show
 the  density distribution $d\Gamma/(d\alpha_1 d\alpha_2)$ with  an ambiguity of $(\alpha_1, \alpha_2)$ 
kinematical configuration. Indeed, the medians of the 
triangle divide 
the allowed $(\alpha_2, \alpha_1)$-plane into six equivalent density 
regions.


Integration over remaining variables $\alpha_1,\alpha_2$ gives us the total splitting width,
\begin{equation}
\label{FinalGamma}
\Gamma_{\gamma\to 3\gamma}\ \simeq\ \,1.2\cdot 10^3\left(\frac{2\alpha^2}{45}\right)^2\frac{1}{2^8\,3!\,\pi^4}\frac{E_{\gamma}^{19}}{m_e^8M^{10}_{LV}} \ \ \simeq\ 5\cdot 10^{-14}\;\frac{E_{\gamma}^{19}}{m_e^8M^{10}_{LV}}.
\end{equation} 
In fact, the parametric dependence of the splitting width (\ref{FinalGamma}) 
coincides with the estimation of \cite{Gelmini:2005gy}, see (\ref{Gelmini}).
Thus, we determine the value of $f$ from (\ref{Gelmini}), which is\footnote{The factor $4\pi^4$ is included into the definition of $f$ in \cite{Gelmini:2005gy}.}  
equal to $f\simeq 3\cdot 10^6$. The numerical factor in (\ref{FinalGamma}) is several orders of magnitude larger than  that estimated in (\ref{Gelmini}). It can be attributed to the large symmetry factor from the matrix element. 

We must verify that the splitting process could be registered in 
observations i.e. that, the most energetic decay products would be 
distinguished from the initial photon which has not undergone
splitting. For this reason we calculate the probability density of 
splitting as the function of maximal photon energy. Thus, we integrate the 
decay width density (see left panel of Fig.~\ref{Density}) over $\alpha_2$. In order to get rid of 
permutation ambiguity, we fix the hierarchy  $\alpha_1 > \alpha_2 > \alpha_3=1-\alpha_1-\alpha_2$, denoting first outgoing photon $k_1$ 
 as the most 
energetic, $\alpha_1\equiv \alpha_{max}$.
This region corresponds to the dark shaded triangle in the left
panel of~Fig.~\ref{Density}.
 The result of numerical integrations are presented in the right panel of
Fig.~\ref{Density}. Note 
that  the probability distribution peaks when the final photon of 
maximal energy carries away one half of the initial 
photon energy, $\alpha_1=0.5$. 
The maximum shifted from $1/3$ to $0.5$. The explanation is the following. There are large number of configurations connected with $\alpha_1=0.5$ but only one configuration for $\alpha_{max}=0.33$. The configuration with tiny energy losses (two soft photons) is still strongly suppressed. 



\section{The constraint from Crab nebula spectrum}

Let us discuss the possibility of splitting for ultra-high-energy photons from astrophysical sources.  
Inverting the width (\ref{FinalGamma}), we obtain the photon mean free path associated with the splitting process,
\begin{equation}
\label{Distance_L}
\langle L \rangle_{\gamma \to 3\gamma}  \simeq 16 \times \left(\frac{M_{LV}}{10^{14}\,\mbox{GeV}}\right)^{10}\, \left( \frac{E_\gamma}{40\, \mbox{TeV}}\right)^{-19}\  \mbox{Mpc}.
\end{equation} 
For the typical photon energy  $\simeq 40$ TeV  and for 
$M_{LV}\simeq 10^{14}$~GeV the photon mean free path is of
 extragalactic scales. 
For a given distance to the source $L_{source}$  there is a cut-off photon energy $E_{cut-off}$, determined by $L  \left(\gamma \to 3\gamma\right) \simeq L_{source}$: photons with larger energy decay via splitting.
The estimation for the constraint on $M_{LV}$  
straightforwardly follows from (\ref{Distance_L}), 
\begin{equation}
\label{Splitting General Bound}
M_{LV} >  \left( \frac{E_\gamma}{40\,\mbox{TeV}}\right)^{1.9} \, \left( \frac{L_{source}}{16\,\mbox{Mpc}}\right)^{0.1} \ \times \; 10^{14} \;\,\mbox{GeV}.
\end{equation}
This bound is applicable if $m_{\gamma, \, eff} \ll m_e$, or 
equivalently
\begin{equation}
\label{condition}
M_{LV} \, \gg \, \frac{E_\gamma^2}{m_e} \; \simeq \; \left( \frac{E_\gamma}{40\,\mbox{TeV}}\right)^2 \ \times \  3\cdot 10^{12}\;\,\mbox{GeV}\,.
\end{equation}
One can see that the bound (\ref{Splitting General Bound}) is stronger than 
the condition (\ref{condition}). Furthermore, the bound 
(\ref{Splitting General Bound}) is sensitive  to the initial 
photon energy  $\sim E_{\gamma}^{1.9}$.   On the other hand, this 
limit weakly depends on the source distance
$\sim L_{source}^{0.1}$.
 Thus, more 
energetic but less distant galactic source can 
set better constraint on 
$M_{LV}$ than less energetic but more distant extragalactic source. 
Let us illustrate it numerically for the galactic Crab nebula 
($L_{source} = 2$ kpc, 
maximal photon energy $E_{\gamma, max} = 75$ TeV \cite{Aharonian:2004gb}); 
and the most energetic extragalactic blazar Mrk 501 ($L_{source} = 140$ 
Mpc, $E_{\gamma, max} =20$ TeV \cite{Aharonian:1999vy}). Formula 
(\ref{Splitting General Bound}) reads, 
\begin{align}
\label{Bound Mrk 501}
\mbox{Crab}: \qquad  M_{LV} > 1.3 \cdot 10^{14}\,\mbox{GeV},\\
\label{Bound Crab Estimation}
\mbox{Mrk} 501: \qquad M_{LV} > 3 \cdot 10^{13}\,\mbox{GeV}. 
\end{align}
The estimated bound from the galactic Crab nebula is 4 times better than from the extragalactic source. Let us perform simple statistical analysis in order to make the bound (\ref{Bound Crab Estimation}) more precise. For this reason we analyze Crab data obtained by HEGRA \cite{Aharonian:2004gb}.
 We closely follow the analysis of Ref.~\cite{Rubtsov:2016bea}.



\begin{figure}[h]
\centering
\hspace{-4mm}
\includegraphics[width=0.53\textwidth]{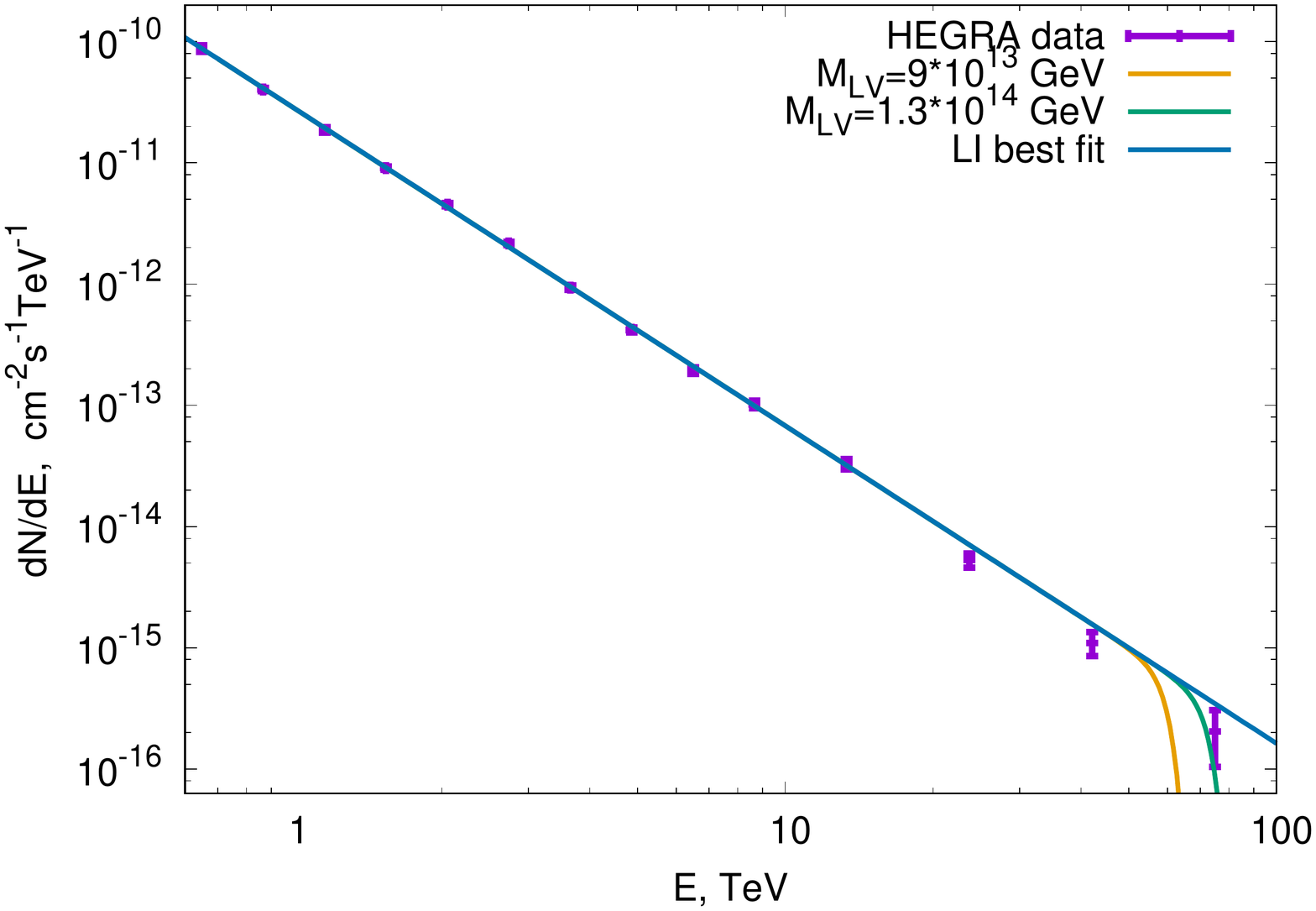}\hspace{-9mm}
\includegraphics[width=0.53\textwidth]{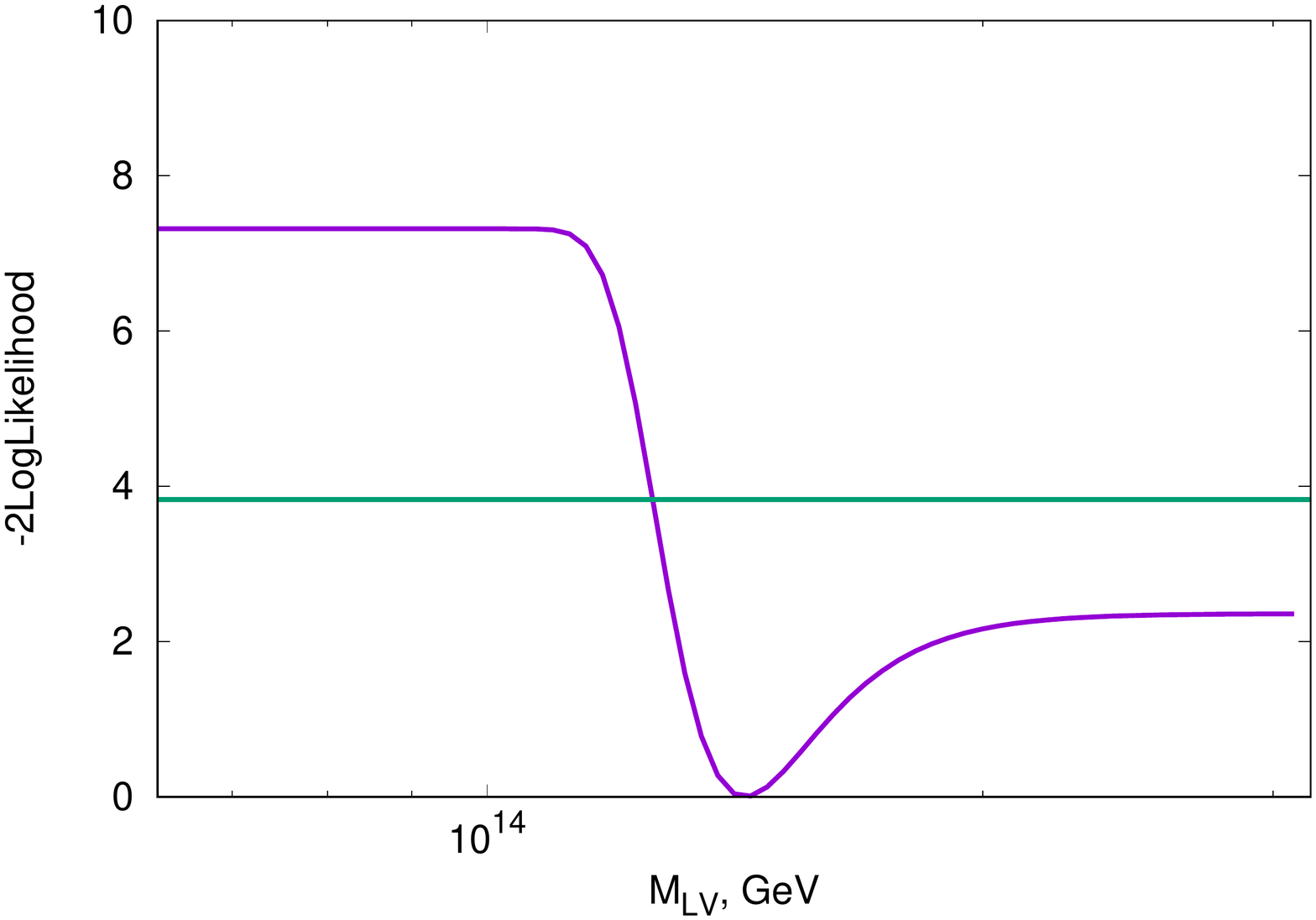}\hspace{-4mm}
\caption{\label{Fig1}
Left panel: photon spectrum of the Crab nebula obtained by collaboration
HEGRA\cite{Aharonian:2004gb}. 
The dashed line corresponds to the best
power-law fit of the spectrum while the dotted lines show the
prediction for the flux under the hypothesis of Lorentz violation 
with a given $M_{LV}$. Right panel: dependence of the likelihood on the LV mass scale $M_{LV}$.
}
\end{figure}

A single splitting process obeys exponential probability distribution. 
Thus, the photon traveling from Crab nebula to Earth  does not undergo the 
splitting with the probability
\begin{equation}
\label{Probability}
P(E) = \mbox{e}^{-L_{CRAB}/\langle L \rangle_{\gamma \to 3\gamma} },
\end{equation}
here $L_{CRAB}$ is the distance to Crab Nebula and the photon mean free path $\langle L \rangle_{\gamma \to 3\gamma}$ is determined by (\ref{Distance_L}).  We expect that the observed Crab Nebula spectrum $d\Phi/dE$ would be corrected by factor of (\ref{Probability}),
\begin{equation}
\label{spectrum}
\left( \frac{d\Phi}{dE}\right)_{LV} = P(E)\cdot \left( \frac{d\Phi}{dE}\right)_{source}.
\end{equation}
The effect of splitting suppression on the HEGRA Crab spectrum \cite{Aharonian:2004gb} is illustrated in Fig.\ref{Fig1}, left panel. We take observational data for photon spectra and use the power-law fit\footnote{A slight steepening may be seen at the end of the spectrum, but its significance is less than $2\,\sigma$.} fixed by data points at energies below 20 TeV, $ \frac{d\Phi}{dE}\sim E^{-2.62}$. We make predictions to the detectable Crab nebula flux under hypothesis of a certain $M_{LV}$.  The break in the highest-energy tail at energy $E_{cut-off}$ of spectrum is clearly seen at Fig.\ref{Fig1}. One can expect also an excess in the spectrum, associated with final products of splitting, located near energy $E_{cut-off}/2$. However, for spectra which decrease stronger than  $\frac{d\Phi}{dE} \propto E^{-2}$ the flux of splitting products at energy $E_{cut-off}/2$ would be less than the flux from the source of that energy. Thus, for Crab nebula, whose spectrum decreases even faster the excess would be negligible small.

We test the family of LV hypotheses parametrized by $M_{LV}$ with HEGRA data\cite{Aharonian:2004gb} for the last bin of Crab nebula spectrum, centered at $75$ TeV. For this reason we apply likelihood ratio method \cite{Li:1983fv,Rolke:2004mj}. More precisely, we test the expectation value for signal events under the hypothesis of LV in the last energy bin with the observed value for signal events, marginalizing over background with unknown expectation value.
 Details of the analysis are described in \cite{Rubtsov:2016bea}. The obtained likelihood profile is shown in Fig. 2, right panel.  
 From it one reads the constraint 
\begin{align}
\label{hegraboundM}
&M_{LV}>1.3\times 10^{14}\,\mbox{GeV}\qquad
&\text{at $95\%$ CL,}
\end{align}
or, in the parametrization of \cite{Kostelecky:2009zp},
\begin{equation}
c^{(6)}_{(I)00} > -1.1\,\times \,10^{-28}\;\ \mbox{GeV}^{-2}\qquad\qquad\qquad\ 
\text{at $95\%$ CL}.
\end{equation}
This constraint is almost two order of magnitude better than obtained from the absence of photon decay (\ref{decayrecent}).

\section{Discussion}
\label{Conclusion}

We have calculated the width of the photon splitting process $\gamma \to 3\gamma$, in LV QED with superluminal photons. We have shown that the splitting process is the leading LV phenomenon in superluminal case that would lead to a  to cut-off in spectra for galactic sources. Using experimental data for Crab spectrum, obtained by HEGRA, we obtained $95\%$ CL lower bounds on the LV mass scale for superluminal dispersion relation for photons. This bound improves previous constraints by more than an order of magnitude.

The ability to constrain $M_{LV}$ from splitting process grows with photon energy as $E_\gamma^{1.9}$. Thus, possible detection of photons with energy exceeding $75$ TeV from a source will set stronger constraints. Crab nebula photon spectrum, according to different theoretical models, continues up to $100$-$400$ TeV \cite{Meyer:2010tta}. At these energies, it would be detected by HAWC\cite{HAWC}, CTA\cite{cta, ctamonte}, Carpet-2\cite{Dzhappuev:2015hxl}, TAIGA\cite{Budnev:2016btu}, LHAASO\cite{DiSciascio:2016rgi}. Detection of photons from Crab (or other galactic sources) in these energy range may increase the bound (\ref{hegraboundM}) up to the level $M_{LV} > 5 \times 10^{14}\, \mbox{GeV}$.

Increasing photon energy significantly more, we can speculate about possible splitting of ultra-high-energy photons in energy range $10^{18} - 10^{19}\,\mbox{eV}$ which can be expected as one of the final products of the Greizen-Zatsepin-Kuzmin (GZK) process \cite{Greisen:1966jv,Zatsepin:1966jv} -- pion production by proton scattering on cosmic microwave background. These GZK, or cosmogenic photons, also can scatter on CMB or radio background. However, a part of the GZK photon flux can reach the Earth. Pierre Auger Observatory or Telescope Array experiments are able to detect predicted flux of GZK photons in near future \cite{Gelmini:2005wu, AlvesBatista:2018zui}, but there are no detected signal yet \cite{Abbasi:2018ywn, Homola:2018gre}.

It is known that  GZK photons would immediately decay to $e^+e^-$ pair in presence of LV of superluminal type \cite{Galaverni:2008yj}. Furthermore, possible observation of a photon with energy $\sim 10^{19}$ eV will set a very strong transplankian constraint $M_{LV} > 10^{23}$ GeV \cite{Galaverni:2008yj}. However, the splitting decay channel is still dominating at this energies. It follows from (\ref{Splitting General Bound}) that future detection of GZK photons $\sim 10^{19}$ eV from GZK horizon $L_{source} \sim 50$ Mpc, will establish the constraint $M_{LV} > 10^{24}$ GeV. This estimated constraint is an order of magnitude better than the bound from the absence of photon decay to $e^+e^-$ pair. However, in order to make a precise constraint both processes of photon production during GZK process and propagation with scattering on cosmic microwave and radio backgrounds should be taken into account in details. 

We also should point out that for qualitative predictions for splitting at these energies we must take into account LV terms for electrons which will produce extra LV terms in the Euler-Heisenberg Lagrangian\footnote{Extra LV terms may be indtroduced to Euler-Heisenberg Lagrangian directly, see \cite{Kostelecky:2018yfa}.}. Detailed computation is out of scope of this article. We leave this interesting task for future.

\paragraph{Acknowledgements} We are very grateful to Sergey Sibiryakov, Grigory Rubtsov, Mikhail Kuznetsov and Mikhail Vysotsky for helpful discussions.


\end{document}